\documentstyle[12pt,aaspp4,epsf]{article}

\newcommand{\etal}{{\it et al.}}
\newcommand{\chisq}{\mbox{$\chi^{2}$}}
\newcommand{\kms}{km\,s$^{-1}$}

\newcommand{\msun}{\mbox{M$_{\odot}$}}
\newcommand{\mdot}{\mbox{$\dot{M}$}}

\newcommand{\smy}{\mbox{M$_{\odot}$\,yr$^{-1}$}}

\newcommand{\bq}{\begin{equation}}
\newcommand{\eq}{\end{equation}}
\newcommand{\ho}{H$_{2}$O}

\newcommand{\ergs}{\mbox{erg\,s$^{-1}$}}

\newcommand{\bpar}{\mbox{B$_{\parallel}$}}
\newcommand{\bperpp}{\mbox{B$_{\perp}$}}

\newcommand{\az}{\mbox{$A_{F-F^{\prime}}$}}
\newcommand{\dnuz}{\mbox{$\Delta\nu_{z}$}}
\newcommand{\dvz}{\mbox{$\Delta v_{z}$}}

\begin{document}
\vskip 0.6 in
\noindent
 
\title{Polarimetric Observations of the Masers in NGC 4258: An
Upper Limit on the Large-Scale Magnetic Field 0.2 pc from the Central 
Engine}
 
\author{J.R. Herrnstein 
   \footnote{Current Address: National Radio Astronomy Observatory, PO Box O,
    Socorro, NM 87801}, J.M. Moran, L.J. Greenhill}
\affil{Harvard-Smithsonian Center for Astrophysics, Mail Stop 42, 
60 Garden Street, Cambridge, MA 02138}
\author{E.G. Blackman}
\affil{Institute of Astronomy, Madingley Road, Cambridge, CB3 0HA}
\author{and P.J. Diamond}
\affil{National Radio Astronomy Observatory, PO Box 0, Socorro, NM 87801}
  
\begin{abstract}
We report VLA $1\sigma$ upper limits of 1.5\% and 3\% on the intrinsic circular 
and linear fractional polarizations, respectively, of the water vapor maser 
emission 0.2 pc 
from the central engine of NGC 4258. A corresponding 0.5\% upper limit on any
Zeeman-splitting-induced circular polarization translates to a 1$\sigma$ upper 
limit on the parallel, or toroidal, component of the magnetic field of 300 mG. 
Assuming magnetic and thermal pressure balance in the disk, this magnetic field 
upper limit corresponds to a model-dependent estimate of the accretion rate 
through the molecular disk of $10^{-1.9}\alpha$\,\smy\ for the case where the magnetic 
field lies along the line of sight.
\end{abstract}


\section{Introduction}
\label{s:mag.1}

Active galactic nuclei (AGN) are thought to be powered by accretion 
onto a central, supermassive black hole (e.g. Rees 1984) and it is 
probable that magnetic 
fields play an important role in this process.  They are often 
invoked as a potential source of viscosity and therefore dissipation
and angular momentum transport in the accretion flow (c.f. Balbus \& Hawley 
1991). In addition, large-scale magnetic fields may be important in the energetics, 
collimation, and confinement of jets and broad line regions (Blandford 
\& Payne 1982; Emmering, Blandford \& Shlosman 1992).  Finally, the 
typical broad-band AGN spectrum is not adequately fit by a simple 
blackbody law, and many of the proposed non-thermal radiation processes 
require a substantial magnetic field.   

Magnetic field strengths can be estimated from radio observations in a variety 
of ways. For example, if the source size can be estimated, then a determination 
of the synchrotron turnover frequency yields an estimate for the field strength. 
Alternatively, if the gas density can be estimated, then Faraday rotation 
measurements can also provide reasonably accurate field strengths. But the
most direct method for estimating magnetic field strengths is via 
observations of the Zeeman splitting of spectral lines of various atomic and 
molecular species in regions of high density neutral gas.  Unfortunately, for 
gas in  thermal equilibrium it is impossible to apply this technique to AGN, 
as the interferometers needed to resolve the central engines of AGN have 
insufficient sensitivity to study the low surface brightness neutral gas 
emission where the Zeeman splitting occurs.  However, the extreme non-thermal 
equilibrium conditions characteristic of extragalactic masers offer an 
unusual opportunity to 
measure AGN magnetic fields. To date approximately 21 \ho\ 
megamasers and 50 OH megamasers have been detected in the nuclei of nearby 
AGN (e.g. Braatz, Wilson \& Henkel 1997). They are often several orders of 
magnitude more luminous than Galactic masers, and many are bright enough 
to be studied with VLBI.

NGC 4258 is a mildly active Seyfert 2 
galaxy with a 2--10 keV X-ray luminosity of $4\times10^{40}$ \ergs\ and an 
obscuring column of $1.5\times10^{23}$\,cm$^{-2}$ (Makishima \etal\ 1994).  
Nuclear continuum and narrow line emission are seen in 
reflected, polarized optical light (Wilkes \etal\ 1995).  The galaxy harbors 
an \ho\ maser (Claussen, Heiligman, \& Lo 1984), and
VLBA observations reveal that the maser traces an extremely thin, slightly 
warped, nearly edge-on disk in the nucleus of the galaxy 
(Watson \& Wallin 1994; Greenhill \etal\ 1995; Miyoshi \etal\ 1995; 
Moran \etal\ 1995; Herrnstein, Greenhill, \& Moran 1996).  The masers
are 0.13--0.26 pc from the center of mass of the disk, assuming a distance 
of 6.4 Mpc.  The nearly perfect Keplerian rotation curve of the masers requires 
a central mass of $3.5\times10^{7}$ \msun\ within 0.13 pc. 
The velocity centroid of 
the disk is consistent with the optical systemic velocity of the galaxy and 
the rotation axis of the disk is well-aligned with a 100-pc-scale radio jet 
(Cecil, Wilson, \& De Pree 1995).  
VLBA observations also reveal subparsec-scale jet emission along the
disk axis (Herrnstein \etal\ 1997).  Deguchi, Nakai, \& Barvainis 
(1995) report an upper limit on the linear polarization of about 20\%
for the highly Doppler-shifted masers and about 15\% for those 
features near the galaxy's systemic 
velocity.

Here, we report an attempt to measure the Zeeman splitting of  
an isolated, strong maser feature in the sub-parsec-scale disk of 
NGC 4258 using VLA 
polarimetric observations. The observations are discussed in 
Section~\ref{s:mag.2}, and the results are interpreted in 
Section~\ref{s:mag.3}. Section~\ref{s:mag.4} considers the 
implications of the measurements.

\section{Observations}
\label{s:mag.2}

\begin{figure}[p]          
\plotone{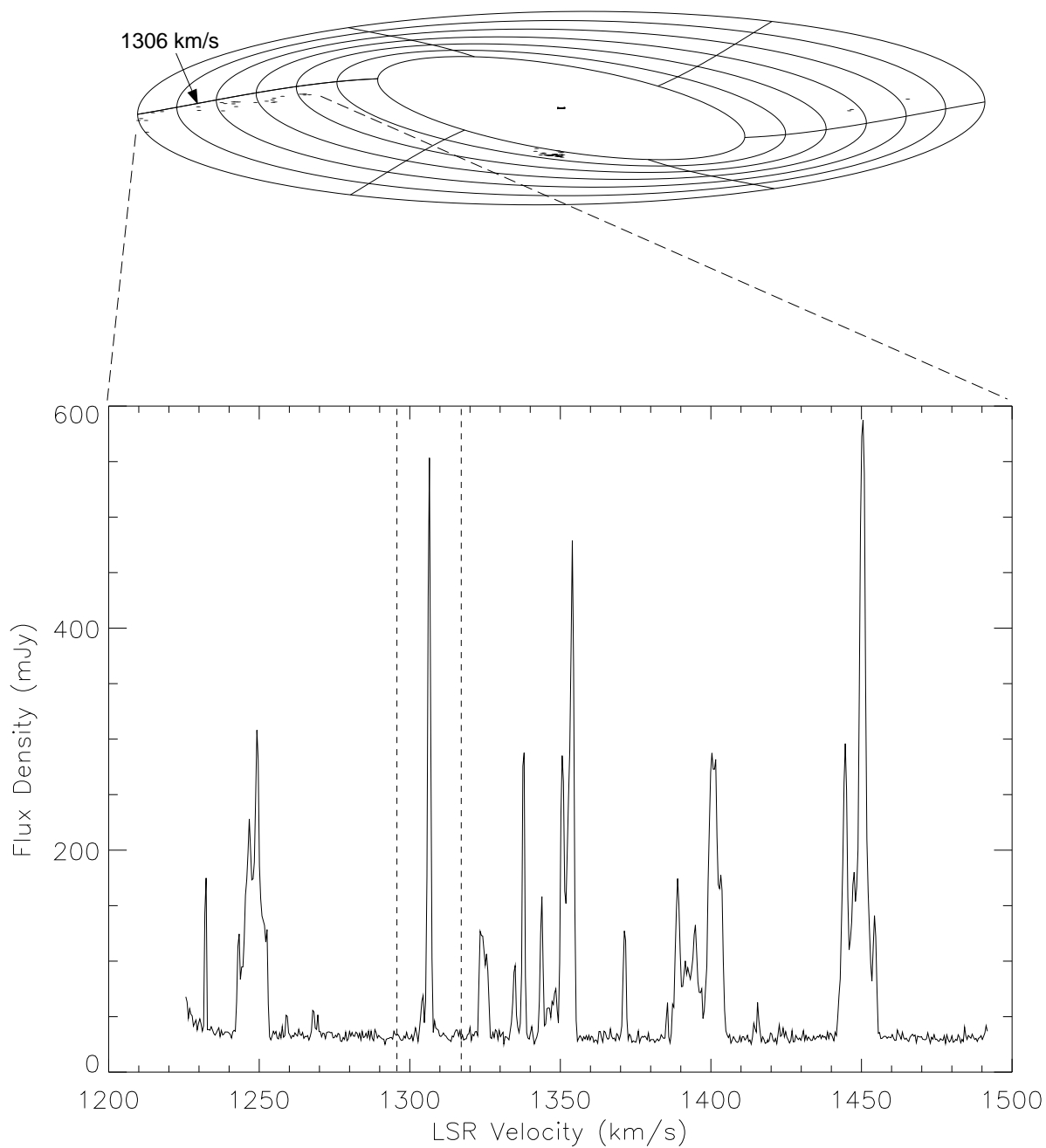}
\caption {{\it Top:} The best-fitting warped disk model in NGC 4258 (Herrnstein 1997), 
with the maser features superposed. The disk has been tipped down 10$^{\circ}$ 
from its true orientation to make the masers more visible. {\it Bottom:} Spectrum 
of the red-shifted maser features. The data were taken with the VLBA on 29 May, 1995 
(Herrnstein 1997). 
The frequency range observed for the current work is indicated by the dashed vertical 
lines.}
\label{fg:mag1}
\end{figure}

The positions and line-of-sight (LOS) velocities of the masers in NGC 4258 define a nearly edge-on,
slightly warped disk, which is shown in the top portion of Figure~1. The figure 
also includes a VLBA spectrum of the red-shifted maser emission.  
The dashed vertical lines in this spectrum show the frequency range observed in the 
current observation. We chose to observe the feature 0.2\,pc from the center of mass 
with a LOS velocity of 1306 \kms\ because it is 
fairly isolated and has remained strong since it was first 
detected by Nakai, Inoue, \& Miyoshi (1993). Because the feature lies approximately 
along the intersection of the disk with the plane of the sky and because the
disk is inclined to the LOS by $\la8^{\circ}$, all subsequent 
references to \bpar\ correspond to any toroidal fields that may be present in 
the disk, while references to the perpendicular component of the field (\bperpp) 
concern either vertical or radial fields. 

The 1306 \kms\ feature was observed for 12 hours on 1995 October 20/21 with 
the B-configuration VLA of the NRAO. \footnote{The National Radio Astronomy 
Observatory is a facility of the National Science Foundation operated under  
cooperative agreement by Associated Universities, Inc.} The skies were clear 
for the first 8 hours, but 
became progressively overcast during the final 4 hours.  A bandwidth of 1.56 
MHz was observed and correlated in full polarization mode with 128 channels
and uniform weighting, and the resulting 0.19\,\kms\ velocity resolution provided 
about six channels 
across the maser line.  Bandpass, amplitude, and phase calibration were 
performed in AIPS using standard techniques.  We were able to self-calibrate 
the 1306 \kms\ feature for the first 8 hours of the experiment.  Linear 
polarization calibration was performed in two steps. First, a strong, 
source (OJ287) was observed over a broad range of parallactic 
angle in order to quantify any discrepancies in the polarization responses of 
the antenna feeds. Second, a strongly polarized source with known polarization 
angle (3C286) was observed to determine the intrinsic phase difference between 
the left and right channels. After imaging, the amplitude calibration was 
refined by fitting Gaussians to the 1306 \kms\ feature and renormalizing the 
left and right circular polarizations (LCP and RCP) to yield equal integrated
line strengths.  This procedure removes any intrinsic circular 
polarization from the data and makes possible the detection of any 
Zeeman-splitting-induced
circular polarization. The renormalization factor was about 0.99, which
suggests there is no intrinsic circularly polarized 
emission (with a spectral profile similar to that of the Stokes I profile) 
to a limit of about 1 percent of the total flux in the line.

\begin{figure}[htbp]          
\plotone{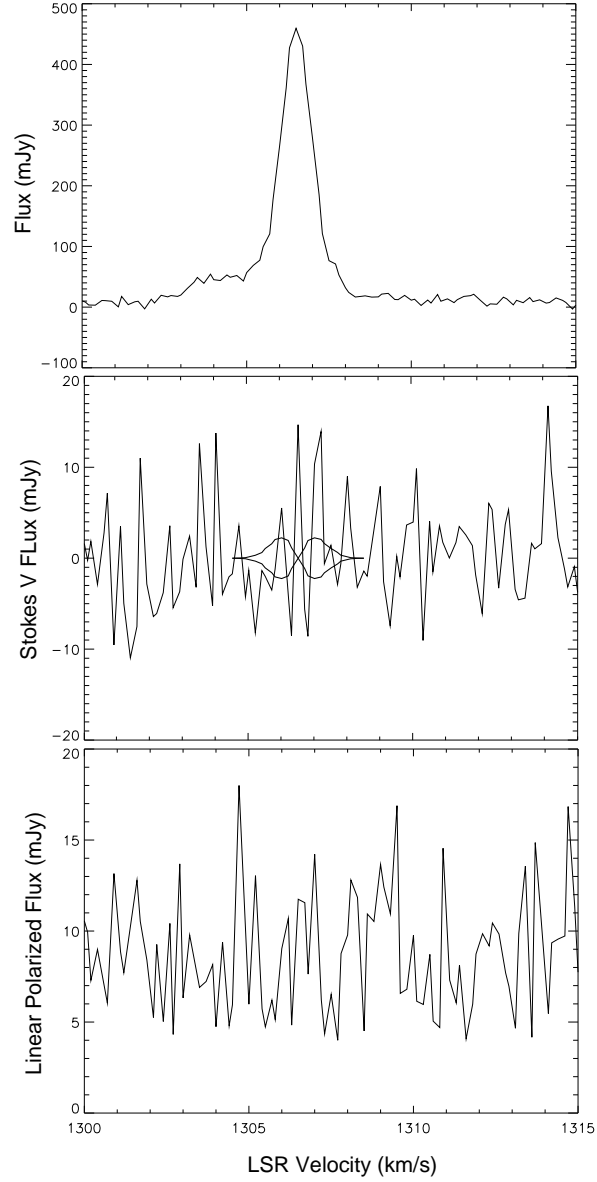}
\caption {VLA polarimetry of the NGC 4258 maser feature at 1306\,\kms.  
{\it Top:} Stokes I spectrum. {\it Middle:} Stokes V spectrum, with the 
$\pm1\sigma$ Zeeman S-spectra superposed. {\it Bottom:} Total linear 
polarization spectrum.} 
\label{fg:mag2}
\end{figure}

The results are summarized in Figure~2. The top panel shows the full 128-channel 
Stokes I ($\equiv 1/2[\mbox{LCP}+\mbox{RCP}]$) synthesized spectrum, with a thermal 
noise of about 6 mJy 
per channel.  Comparison with Figure~1 shows that the flux density of the 1306 \kms\ 
feature has remained steady at approximately 0.5 Jy over the 6-month time baseline. 
The best fitting Gaussian is 
centered at 1306.4 \kms\ (LSR, radio definition) and has a line width, $\sigma_{v}$, 
of 0.49\,\kms\ 
(FWHM, $\Delta v$, of 1.15 \kms). We note that the actual deconvolved FWHM is
about 1.13\,\kms. The Stokes V ($\equiv 1/2[\mbox{LCP}-\mbox{RCP}]$) 
spectrum is shown in 
the middle panel of Figure~2, while the bottom panel shows the total linear polarization 
spectrum (Stokes $\sqrt{Q^{2}+U^{2}}$). 
{\it There is no evidence for any intrinsic polarization 
in the 1306 \kms\ maser feature, and we report $1\sigma$ upper limits 
of 1.5\% and 3\% on the fractional circular (V/I) and linear polarizations, 
respectively.}  

\section{Interpretation}
\label{s:mag.3}

An external magnetic field, $\vec{B}$, induces a precession in the magnetic 
moment of the water molecule about $\hat B$.  The energy of this precession 
generates magnetic sublevels ($M_{F}$) by lifting the degeneracy within the 
hyperfine states of water. Zeeman splitting arises from transitions between 
these sublevels, according to the selection rule $\Delta M_{F}= 0, \pm 1$. The 
$\Delta M_{F}=\pm 1$ transitions give rise to line emission asymmetrically 
shifted in frequency from the hyperfine transition rest frequency, and
possessing counter-rotating circular polarizations in the plane perpendicular to 
$\hat B$.  Because water is non-paramagnetic, $\vec B$  couples primarily to 
the nuclear magnetic dipole moment of the molecule, and the splitting of the left 
and right circular polarizations (\dnuz) is only about 10 Hz mG$^{-1}$ (c.f.
Feibig \& G\"{u}sten 1989; hereafter FG). Thus, for mG-level magnetic fields 
the Zeeman splitting is $10^{-3}$--$10^{-4}$ times narrower than typical 
water maser linewidths.  For such fractional Zeeman splitting, the 
Stokes V spectrum is well-approximated as the first derivative of the total 
power spectrum, I, times the splitting width:
\bq
V(v) \simeq \frac{dI}{dv}\dvz = \frac{I_{o}\dvz}{\sigma^{2}_{v}} v e^{-\frac{v^{2}}{2\sigma^{2}_{v}}},
\label{eq:bfirst}
\eq
where we have modeled the line as a Gaussian with amplitude $I_{o}$, and where \dvz\
is the Zeeman splitting in units of velocity and $v$ is the offset from line center
in velocity. Equation~\ref{eq:bfirst} is an
antisymmetric sinusoid, or ``S-curve'' with extrema of $\pm e^{-1/2}I_{o}\dvz/\sigma_{v}$
at $v=\pm\sigma_{v}$.  Note that the spectrum need only be sampled at a fraction 
of $\sigma_{v}$, as opposed to a fraction of \dvz, in order to detect the Zeeman 
splitting.

In practice, the Zeeman pattern of \ho\ is more complex since in the presence of
a magnetic field the 7--6, 6--5, and 
5--4 hyperfine transitions that dominate the water maser spectrum are themselves 
comprised of 13, 11, and 9 different lines in each polarization, each with its 
own characteristic \dvz. FG have modelled the  
theoretical \ho\ Zeeman pattern, and they find that the S-curve extrema ($V_{max}$) 
are approximately given by
\bq
\frac{V_{max}}{I_{o}}\simeq\az\frac{\bpar \mbox{~~[G]}}{\Delta v\mbox{~~[\kms]}},
\label{eq:fg1}
\eq
where \az\ equals 0.0133, 0.0083, and 0.001\,\kms\,G$^{-1}$ for the 7-6, 6-5, and 
5-4 transitions, respectively.  Polarimetric observations of \ho\ masers interpreted 
with equation~\ref{eq:bfirst} and normalized according to equation~\ref{eq:fg1} have 
been used to infer interstellar magnetic fields of up to 100 mG in galactic star 
formation regions (FG; Zhao, Goss, \& Diamond 1992).  FG find that the magnetic fields
measured using Zeeman
splitting correlate well with density in molecular clouds over several 
orders of magnitude.
This is consistent with expectations based on flux freezing arguments, and provides 
circumstantial support for the validity of the Zeeman technique.

Equations~\ref{eq:bfirst} and \ref{eq:fg1} neglect the unusual radiative transfer 
properties and nonthermal nature of masers. Nedoluha \& Watson (1992) explore 
the validity of the FG S-curve formulation for fractional Zeeman splitting in light 
of the full transfer equations and considering all hyperfine transitions.  Their 
numerical simulations verify the standard Zeeman splitting methodology in the regime 
of mild maser saturation and suggest that $\az=0.020$\,\kms\,G$^{-1}$ accurately accounts 
for all hyperfine transitions taken together.  For extreme saturation, effects such 
as line re-broadening and non-Maxwellian particle velocities may become significant, 
limiting the precision of the  magnetic field estimates to factors of a few.  
In addition, as the stimulated emission rate, $R$, approaches \dnuz\ for highly 
saturated masers, strong ``false'' circular polarizations can be induced that are 
similar in form to the Zeeman pattern, and therefore conducive to {\it overestimates} 
of the magnetic field (Nedoluha \& Watson 1990). 

The results of the numerical simulations emphasize that the degree of saturation is
important in interpreting maser polarization observations.  For a cylindrical maser 
(c.f. Reid and Moran 1988)
\bq
  R\simeq8\left[\frac{S_{\nu}}{1\mbox{~Jy}}\right]\left[\frac{l}{10^{-2}\mbox{~pc}}\right]^{-2}\left[\frac{d}{6.4\mbox{~Mpc}}\right]^{2}\mbox{~~~s$^{-1}$}.
\label{eq:sat1}
\eq
where $S_{\nu}$ and $l$ are the flux density and length of the maser, respectively, 
and $d$ is the distance to the source.  A maser becomes saturated when $R$ surpasses 
$\Gamma$, the relaxation rate from the maser levels, which is usually assumed to be  
$\la1$\,s$^{-1}$ for the relevant levels of the water molecule.  In NGC 4258, both 
velocity coherence and beam angle arguments suggest $l\la0.002$\,pc for the 
high-velocity masers (Moran \etal\ 1995).  Therefore, $R\ga10^{2}$\,s$^{-1}$ for the 
$\sim0.5$\,Jy feature at 1306\,\kms, and the maser is probably highly saturated.  We 
note that the 1.1 \kms\ FWHM of the line is consistent with that of a re-broadened, 
highly saturated maser at $\sim300$\,K (Nedoluha \& Watson 1991).

We derive an upper limit on \bpar\ from the observed Stokes V spectrum using 
equations~\ref{eq:bfirst} and \ref{eq:fg1} with $\az=0.020$\,\kms\,G$^{-1}$ 
and $\Delta v=1.15$\,\kms. 
{\it A \chisq\ minimization procedure in which \bpar\ is the only free parameter 
indicates that $V_{max} \la 2.2$ \,mJy and $\vert\bpar\vert \la 300$\,mG 
($1\sigma$).  Here, \bpar\ represents any toroidal component of the magnetic 
field at 0.2\,pc radius that is not significantly tangled across the maser.} 
The $\pm300$\,mG Zeeman S-curve corresponding to the upper limit on \bpar\ are 
included in the middle panel of Figure~2.  If the maser line consists of only the 
7--6 hyperfine component, as is commonly assumed,  then the field strength 
limit would be about 50\% higher.  Finally, we note that if the maser is 
indeed highly saturated, the upper limit is uncertain by factors of a few. 

Interpreting the linear polarization upper limit is more difficult, as there is
substantial uncertainty surrounding the theory of linear polarization in masers 
exposed to an external magnetic field (c.f. Deguchi \& Watson 1990; Elitzur 1996).  
In general, the degree of linear polarization will probably depend quite strongly 
on the saturation state of the maser (Goldreich, Keeley, \& Kwan 1973; 
Nedoluha \& Watson 1992), and this injects considerable uncertainty into field strength
estimates based on linear polarization observations. Furthermore, it is difficult
to quantify the degree to which the linear polarization observations have been affected 
by Faraday depolarization within the maser itself, since the ionization fraction 
and number density ($n$) in the molecular layer of the disk are both rather uncertain.  
For $n\sim10^{10}$\,cm$^{-3}$, a magnetic field of 300\,mG, and a maser length of 
0.002 pc, a fractional ionization as low as $10^{-8}$ would lead to significant
Faraday depolarization (Thompson, Moran, \& Swenson 1986).  We note that a recent 
analysis of 
circumstellar SiO masers suggests that Faraday depolarization is not important in 
such systems (Wallin \& Watson 1997). However, in the present analysis we will 
refrain from translating the 3\% upper limit on the linear polarization into an upper 
limit on the magnetic 
field.

\section{Discussion and Future Prospects}
\label{s:mag.4}

The upper limit on the magnetic field at 0.2 pc radius can be compared to a second upper 
limit relating to pressure balance in the disk. Moran \etal\ (1995) use the scatter 
in the vertical component of the systemic masers to place an upper limit of 0.0003\,pc 
on the scale height, $H$, of the maser layer.  If the disk is supported by magnetic pressure 
and is in hydrostatic equilibrium then $H/r = v_{A}/v_{\phi}$, where $r$ is the radius, 
$v_{A}$ is the Alfv$\acute{\mbox{e}}$n velocity ($v_{A}=B/\sqrt{4\pi\rho}$), and $v_{\phi}$ 
is the Keplerian rotation velocity. $\rho$ is the density in the maser layer, and the 
Alfv$\acute{\mbox{e}}$n velocity in question relates to any magnetic field geometry 
capable of supplying 
pressure in the vertical direction: tangled, radial, and toroidal fields all suffice. 
We require $n \la 10^{10}$ cm$^{-3}$ in order to support maser action and conclude that 
$B \la 80$ mG in the maser layer at $r=0.2$\,pc, the equality holding for a magnetically 
dominated disk characterized by the quoted upper limits for thickness and density.  
The 300 mG upper limit on \bpar\ provided by the Zeeman analysis does not preclude a 
magnetically supported disk. It does, however, provide a more robust upper limit on 
the field strength than do hydrostatic equilibrium considerations, since the masers
could in principle be sampling a thin layer in a much thicker disk.

NGC 4258 is an exceptional laboratory for the study of AGN accretion processes. The 
nuclear maser reveals details about the kinematics and structure of the accretion disk 
on sub-parsec scales and permits the determination of the central mass with great 
precision. In addition to being constrained by these observations, 
efforts to construct self-consistent models of the accretion in NGC 4258 must 
necessarily address the fact that the central engine luminosity ($L$) is extremely 
sub-Eddington: ($L\sim10^{42\pm1}~\ergs=10^{-3.6\pm1}L_{E}$; Herrnstein \etal\ 1998).  
Neufeld \& Maloney 
(1995) demonstrate that for an accretion rate, $\mdot$, of approximately 
$10^{-4.1}\alpha$\,\smy, the molecular disk becomes atomic throughout for radii greater 
than about 0.26 pc, thereby providing a 
physical explanation for the observed outer limit of the maser emission. Here, 
$\alpha$ is the standard Shakura-Sunyaev parameterization of the kinematic viscosity.
This accretion 
rate corresponds to a radiative efficiency of $10^{-0.6\pm1}\alpha^{-1}$ and suggests 
that NGC 4258 is powered by a highly efficient, but mass-starved, optically thick, 
geometrically thin accretion disk (Shakura \& Sunyaev 1973).  Alternatively, Lasota 
\etal\ (1996) have proposed that NGC 4258 is powered by a central geometrically thick, 
optically thin advection-dominated accretion flow (ADAF; Narayan \&  Yi 1995a\&b).  
In this scenario, the vast majority of viscously dissipated energy is deposited into 
an extremely hot ($\sim10^{12}$\,K), largely non-radiative proton plasma, and the 
low luminosity of the system is due to low radiative efficiencies as opposed to mass 
starvation of the central black hole.  In the ADAF models the broad-band spectrum, from 
radio to hard X-ray, is the result of various nonthermal processes in a relatively cool 
($\sim10^{9.5}$\,K) electron plasma.  Lasota \etal\ (1996) fit the optical and X-ray 
NGC 4258 
emission with an $\mdot\simeq10^{-1.9}\alpha$\,\smy\ ADAF.  More recent calculations 
by Narayan (personal communication), suggest $\mdot\simeq10^{-1.7}\alpha$\,\smy, 
corresponding to a radiative efficiency of $10^{-3\pm1}\alpha^{-1}$.  We note that recent 
high-resolution radio continuum observations suggest that if NGC 4258 harbors an 
ADAF, it must be truncated within about 100 Schwarzschild radii (Herrnstein \etal\ 
1998).

\begin{figure}[htb]          
\plotone{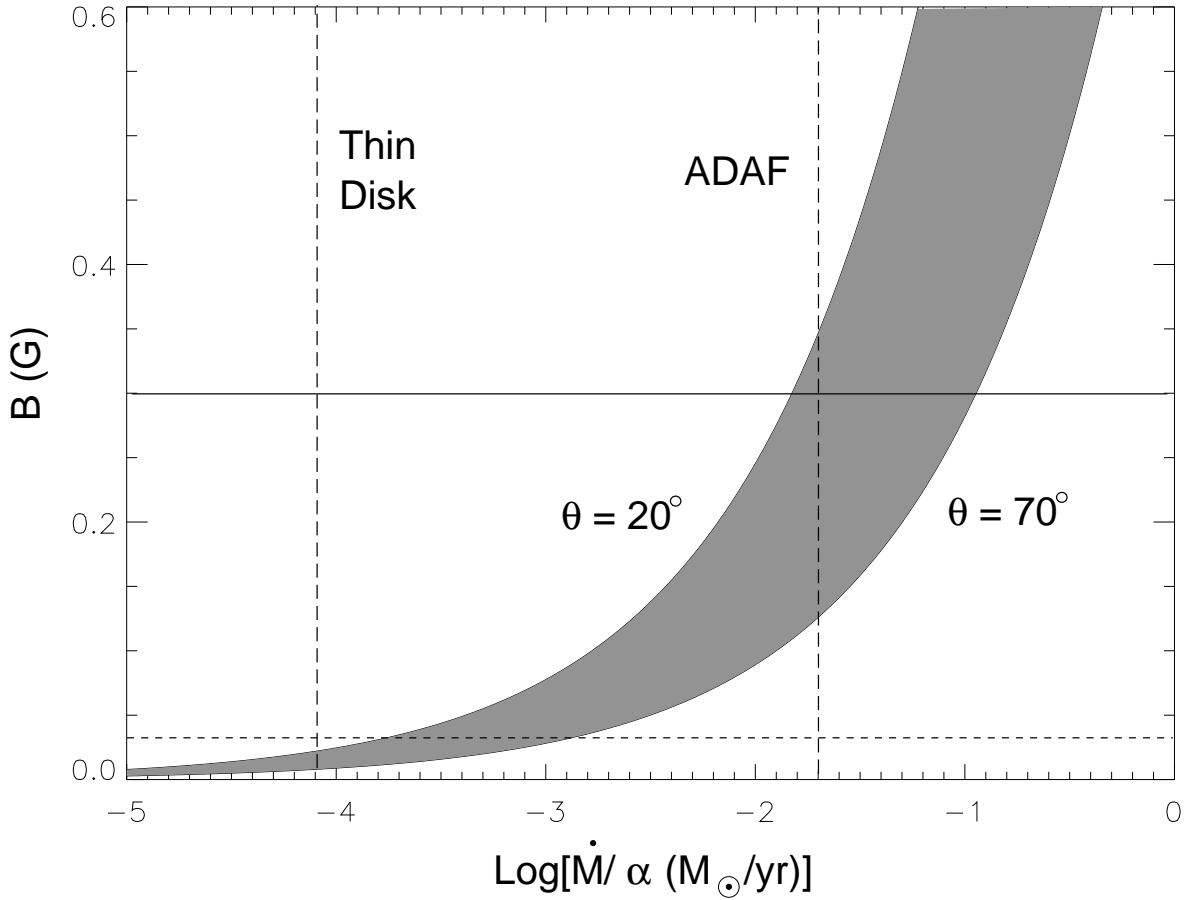}
\caption {The shaded region shows the expected strength of the LOS-component of 
the equipartition magnetic field as a function of $\mdot/\alpha$ at a radius of 
0.2 pc for a temperature of 600\,K and for a range of field orientations. The 
accretion rates predicted 
by the thin-disk and ADAF models are shown (dashed vertical lines), as well as the 
upper limit on \bpar\ from the present experiment (solid horizontal line).  The
dashed horizontal line shows the theoretical magnetic field sensitivity of a proposed
``broad-band'' Zeeman experiment.} 
\label{fg:mag3}
\end{figure}

An observational determination of \mdot\ would be the most unambiguous way to 
discriminate between these two modes of accretion.  Unfortunately, \mdot\ is very 
difficult to measure. The most direct constraint is derived from the extreme 
precision of the maser Keplerian rotation curve, and requires 
$\mdot \la 10^{-1}\alpha$\,\smy (Herrnstein \etal\ 1998).  
The present upper limit on \bpar\ provides an additional, albeit less direct, 
probe of \mdot.  Following Neufeld \& Maloney (1995), we treat the X-ray-heated 
outer disk in NGC 4258 
as a thin isothermal disk in hydrostatic equilibrium, in which the standard 
Shakura-Sunyaev accretion solutions apply.  The midplane pressure, $p_{c}$, is 
then given by (Frank, King, \& Raine 1992)
\bq
p_{c}=\frac{1}{3\sqrt{2\pi^{3}}}\frac{GM\mdot}{\alpha c_{s} r^{3}},
\label{eq:nm1}
\eq
where $M$ is the central mass and $c_{s}$ is the isothermal sound speed.
The magnetic field strength can be related to the accretion rate using 
equation~\ref{eq:nm1} under the assumption of thermal and magnetic pressure
balance in the disk ($\vert B\vert^{2}/8\pi=p_{c}$).  One potential concern 
is that the very low ionization fraction in the molecular disk will 
significantly limit the coupling between the magnetic fields and the gas.  
However, we note that direct exposure to the central X-ray source (as a 
result of the warp), will generate a layer of atomic gas on top of the 
molecular disk with 1--2\% fractional ionization, and that at $r=0.2$\,pc 
the molecular layer is only about 20\% of the total disk thickness (Neufeld, 
Maloney, \& Conger 1994; Herrnstein, Greenhill, \& Moran 1996).  The parallel 
component of the magnetic field is related to $B$ according to $\bpar = uB$, 
where $u=\cos(\theta)$ and $\theta$ is the angle between the LOS and the 
magnetic field at the position of the maser in question. A disk permeated by 
large-scale vertical or radial fields has $u\simeq0$ along the midline, 
while a disk dominated by large-scale toroidal fields has $u\simeq1$.
{\it Adopting a temperature of 600\,K in order to accommodate the presence 
of masers in the disk, the present 300\,mG upper limit on $\vert\bpar\vert$ 
together with equation~\ref{eq:nm1} implies $\mdot\la10^{-1.9}\alpha u^{-2}$\,\smy.}  
Figure~3 shows the theoretical LOS magnetic field strength 
as a function of \mdot\ for $20^{\circ}<\theta<70^{\circ}$ and illustrates 
that because of ambiguities in the magnetic field geometry, the present 
observations cannot distinguish between the ADAF and thin-disk models.  
We emphasize that this analysis implicitly assumes that a substantial component 
of the equipartition magnetic field is constant across the cross-section of 
the maser feature at 1306\,\kms, which is thought to have a linear scale of 
$\sim10^{15}$\,cm (Moran \etal\ 1995).  Furthermore, the theoretical field 
estimates are only as accurate as the assumption of pressure balance in the 
disk. It is at least in principle possible to construct either thermally or 
magnetically dominated disk models.

While the present VLA observations are not sufficiently sensitive to 
discriminate between the low and high \mdot\ models, it may eventually 
be possible to do this by extending the Zeeman S-curve formalism to search 
for coherent, ``broad-band'' Zeeman splitting across the systemic masers, 
which arise on the near side of the disk (see Figure~1).  In such a 
broad-band polarimetry experiment, one could search for a coherent 
frequency offset between the LCP and RCP spectra, by comparing $dI/d\nu$
to the Stokes V spectrum across the entire 100 \kms\ of systemic emission. 
Because 
the systemic emission is 100 times broader and 10 times brighter than the 
feature at 1306 \kms, the magnetic field sensitivity would be greatly 
enhanced.  Simulations of this experiment using actual VLBA spectra predict 
a 1$\sigma$ uncertainty in \bpar\ of 30 mG, which in principle can
discriminate between the low and high \mdot\ models, even accounting
for ambiguities in the field configuration (dashed horizontal line
in Figure~3). The 
broad-band Zeeman measurement would be sensitive to that component of 
the radial magnetic field that is constant across the systemic masers, 
which span about 4$^{\circ}$ in disk azimuth and 
$10^{16}$ cm in radius.


\begin{thebibliography}{}
\bibitem[Balbus \& Hawley 1991]{balbus} Balbus, S. A. \& Hawley, 
J. F. 1991, {\it ApJ}, 376, 214
\bibitem[Blandford and Payne 1982]{blandford2} Blandford, R. D. \& 
Payne, D. G. 1982, {\it MNRAS}, 199, 883
\bibitem[Braatz \etal 1997]{braatz} Braatz, J. A., Wilson, A. S., \&
Henkel, C. 1997, {\it ApJs}, 110, 231 
\bibitem[Cecil \etal 1995]{cecil2yy} Cecil, G., Wilson, A. S., \& 
De Pree 1995, {\it ApJ}, 440, 181
\bibitem[Claussen \etal 1984]{claussen} Claussen, M. J., Heiligman, 
G. M., \& Lo, K. Y. 1984, {\it Nature}, 310, 298
\bibitem[Deguchi and Watson 1990]{deguchi1yy} Deguchi, S. \& Watson, 
W. D. 1990, {\it ApJ}, 354, 649
\bibitem[Deguchi \etal 1995]{deguchi2yy} Deguchi, S., Nakai, N., \&
Barvainis, R. 1995, {\it AJ}, 109, 507
\bibitem[Elitzur 1996]{elitzur3yy} Elitzur M. 1996, {\it ApJ}, 457, 415
\bibitem[Emmering \etal 1992]{emmering} Emmering, R. T., Blandford, 
R. D., \& Shlosman, I. 1992, {\it ApJ}, 385, 460
\bibitem[Fiebig and Gusten 1989]{fiebigyy} Fiebig, D. \&
Gusten, R. 1989, {\it A\&A}, 214, 333 (FG)
\bibitem[Frank \etal 1992]{frank1yy} Frank, J., King, A., \& Raine, 
D. 1992, {\it Accretion Power in Astrophysics, 2$^{nd}$ ed.}, 
Cambridge University Press
\bibitem[Goldreich \etal 1973]{goldreichyy} Goldreich, P., Keeley, 
D. A., \& Kwan, J. Y. 1973, {\it ApJ}, 179, 111
\bibitem[Greenhill \etal 1995]{greenhill1yy} Greenhill, L. G., 
Jiang, R. D., Moran, J. M., Reid, M. J., Lo, K. Y., \& Claussen, M. J. 1995, 
{\it ApJ}, 440, 619
\bibitem[Herrnstein \etal 1998]{herrnstein2yy} Herrnstein, J. R. , Greenhill, L. J., 
Moran, J. M., Diamond, P. J., Inoue, M., Nakai, N., \& Miyoshi, M. 1998, 
{\it ApJ}, in print
\bibitem[Herrnstein \etal 1997]{herrnstein2xy} Herrnstein, J. R. , Moran, J. M., 
Greenhill, L. J., Diamond, P. J., Miyoshi, M., Nakai, N., \& Inoue, M. 1997, 
{\it ApJ}, 475, L17
\bibitem[Herrnstein \etal 1996]{herrnstein1yy} Herrnstein, J. R., 
Greenhill, L. J., \& Moran, J. M. 1996, {\it ApJ}, 468, L17
\bibitem[Lasota \etal 1996]{lassie} Lasota, J.-P., Abramowicz, M. A., Chen, X.,
Krolik, J., Narayan, R., \& Yi, I. 1996, {\it ApJ}, 462, 142
\bibitem[Makishima \etal 1994]{makishima1yy} Makishima, K.,
Fujimoto, R, Ishisaki, Y., Kii, T., Lowenstein, M., Mushotzky, R., 
Serlemitsos, P., Sonobe, T., Tashiro, M., \& Yaqoob, T. 1994, {\it Proc. 
Astron. Soc. Jpn.}, 46, L77  
\bibitem[Miyoshi \etal 1995]{mineyy} Miyoshi, M., Moran, J. M., 
Herrnstein, J. R., Greenhill, L. J., Nakai, N., Diamond, P. J., \& 
Inoue, M. 1995, {\it Nature}, 373, 127
\bibitem[Moran \etal 1995]{moran1yy}  Moran, J. M., Greenhill, L. J.,
Herrnstein, J. R., Diamond, P. J., Miyoshi, M., Nakai, N., \& Inoue, M. 
1995, {\it Proc. Natl. Acad. Sci. USA}, {\bf 92}, 11427
\bibitem[Nakai \etal 1993]{nakai1yy} Nakai, N., Inoue, M., \& 
Miyoshi, M. 1993, {\it Nature}, 361, 45
\bibitem[Narayan \& Yi 1995a]{narly1} Narayan, R. \& Yi, I. 1995a, 
{\it ApJ}, 444, 231 
\bibitem[Narayan \& Yi 1995b]{narly2} Narayan, R. \& Yi, I. 1995b, 
{\it ApJ}, 452, 710 
\bibitem[Nedoluha and Watson 1992]{nedoluha4yy} Nedoluha, G. E. \&
Watson, W. D. 1992, {\it ApJ}, 384, 185
\bibitem[Nedoluha and Watson 1991]{nedoluhadde} Nedoluha, G. E. \&
Watson, W. D. 1991, {\it ApJ}, 367, L63
\bibitem[Nedoluha and Watson 1990]{nedoluhadd} Nedoluha, G. E. \&
Watson, W. D. 1992, {\it ApJ}, 361, L53
\bibitem[Neufeld \& Maloney 1995]{noise} Neufeld, D. A. \& Maloney, P. 
R. 1995, {\it ApJ}, 447, L17
\bibitem[Neufeld \etal\ 1994]{neufeld1} Neufeld, D. A., Maloney, P. R., 
\& Conger, S. 1994, {\it ApJ}, 436, L127
\bibitem[Rees 1984]{rees1yy} Rees, M. J. 1984, {\it Ann. Rev. Astron. 
Astrophys.}, 22, 471
\bibitem[Reid \& Moran 1988]{reid11} Reid, M. J., \& Moran, J. M. 1988, in 
{\it Galactic and Extragalactic Radio Astronomy}, ed. G. L. Verschuur \& K. I. 
Kellermann (Berlin: Springer-Verlag), p. 255
\bibitem[Shakura \& Sunyaev]{sigar} Shakura, N. I. \& Sunyaev, R. A. 
1973, {\it A\&A}, 24, 337 
\bibitem[Thompson, Moran, \& Swenson 1986]{tms} Thompson, A. R., Moran, J. M., 
\& Swenson G. W. 1986, {\it Interferometry and Synthesis in Radio Astronomy},
Wiley Interscience, New York
\bibitem[Wilkes \etal 1995]{wilkes1yy} Wilkes, B. J., Schmidt, G. D.,
Smith, P. S., Mathur, S., \& McLeod, K. K. 1995, {\it ApJ}, 455, L13
\bibitem[Wallin \& Watson 1997]{wally} Wallin, B. K. \& Watson, W. D.
1997, {\it ApJ}, 481, 832
\bibitem[Watson \& Wallin 1994]{watsona} Watson, W. D. \& Wallin, B. K. 1994, 
{\it ApJ}, 432,L35
\bibitem[Zhao \etal 1989]{zhaoyy} Zhao, J., Goss, W. M., \&
Diamond, P. J. in {\it Astrophysical Masers}, Springer, 1992, page 180
\end{thebibliography}
\end{document}